\documentclass[  a4paper,
                 ,jap
                 ,twocolumn
                 ,superscriptaddress
               ]{revtex4}

\usepackage[utf8x]{inputenc}
\usepackage{amsmath}
\usepackage{amsfonts}
\usepackage{amssymb}
\usepackage[english]{babel}
\usepackage{fontenc}
\usepackage{graphicx}
\usepackage{natbib}

\usepackage{hyperref}
\graphicspath{{images/}}
\hypersetup{colorlinks=true,linkcolor=blue,citecolor=blue,pdftitle={Numerical studies of the EUV-induced plasma in argon and hydrogen}}

\newcommand*{\doi}[1]{\href{http://dx.doi.org/\detokenize{#1}}{doi:\detokenize {#1}}}

\begin{document}
\title{Exploring the electron density in plasma induced by EUV radiation: II. Numerical studies in argon and hydrogen}
 \author{D.I.~Astakhov}
 \affiliation{XUV Group, MESA+ Institute for Nanotechnology, University of Twente, P.O. Box 217, 7500 AE Enschede, The Netherlands}
 \author{W.J.~Goedheer}
 \affiliation{FOM Institute DIFFER - Dutch Institute for Fundamental Energy Research, PO Box 6336, 5600HH Eindhoven, the Netherlands, The Netherlands}
 \author{C.J.~Lee}
 \affiliation{XUV Group, MESA+ Institute for Nanotechnology, University of Twente, P.O. Box 217, 7500 AE Enschede, The Netherlands}
 \author{V.V.~Ivanov}
 \author{V.M.~Krivtsun}
 \author{K.N.~Koshelev}
 \affiliation{Institute for Spectroscopy RAS (ISAN), Fizicheskaya 5, Troitsk 142190, Russian Federation}
 \author{D.V.~Lopaev}
 \affiliation{Skobeltsyn Institute of Nuclear Physics, Lomonosov Moscow State University, Leninskie Gory, Moscow 119991, Russian Federation}
 \author{R.M.~van~der~Horst}
 \author{J.~Beckers}
 \affiliation{ Department of Applied Physics, Eindhoven University of Technology, PO Box 513,
5600MB Eindhoven, The Netherlands}
 \author{E.A.~Osorio}
 \affiliation{ ASML The Netherlands B.V., PO Box 324, 5500AH Veldhoven, The Netherlands}
 \author{F.~Bijkerk}
 \affiliation{XUV Group, MESA+ Institute for Nanotechnology, University of Twente, P.O. Box 217, 7500 AE Enschede, The Netherlands}

\begin{abstract}
We used numerical modeling to study the evolution of EUV-induced plasmas in argon and hydrogen. The results of simulations were compared to the electron densities measured by microwave cavity resonance spectroscopy. It was found that the measured electron densities can be used to derive the integral amount of plasma in the cavity. However, in some regimes, the impact of the setup geometry, EUV spectrum, and EUV induced secondary emission should be taken into account. The influence of these parameters on the generated plasma and the measured electron density is discussed.
\end{abstract}

\maketitle
\section{Introduction}
In the semiconductor industry, the photolithography process is used to create patterns on  silicon wafers,  an important step in microchip production. Due to diffraction, the characteristic pattern dimensions depend on the illuminating wavelength.  However, it is becoming increasingly complicated to further reduce printed feature sizes with deep ultraviolet (DUV) light sources. As a result, extreme ultraviolet (EUV) lithography, operating at a wavelength of 13.5~nm, is expected to be used to print integrated circuits with very high resolution features.

For various reasons, a buffer gas is frequently used in both laboratory and industrial EUV light sources. However, even at  low pressure (1~..~30~Pa),  EUV absorption leads to EUV induced plasma formation. There is a strong need for reliable diagnostics of EUV induced plasmas, because the interaction of an EUV induced plasma with the chamber interior and optical elements can lead to various plasma induced processes, such as surface etching, accelerated deposition of overlayers, and oxidation, depending on the exact constituents of the background gas \cite{Madey.2006.surface,Davis.2007.situ}.

It was shown in \cite{Horst.2014.exploring} that it is possible to measure the electron density of an EUV induced plasma,   at  relevant background gas pressures, by measuring the resonant frequency shift of a microwave cavity that contains the plasma. Although this is a highly sensitive method, it  only allows the field average~\cite{Gundermann.2001.microwave} electron density inside the cavity to be determined. 

Nevertheless, it is interesting to know the spatial distribution of plasma in the cavity, the electron temperature and other plasma parameters. Unfortunately,  measuring all the relevant parameters experimentally is difficult  due to the  transient nature of the  EUV induced plasma.  For example,  the insertion of a probe can yield unreliable estimates of plasma parameters for EUV induced plasmas,  since, during and after the EUV pulse, the probe signal is heavily distorted \cite{Velden.2008.radiation,Dolgov.2015.extreme}. 

In this paper, we use a particle-in-cell model of the EUV-induced plasma to determine the spatio-temporal distribution of the electron density, the electron-ion balance, and the influence that some confounding factors may have on the spatio-temporal distribution of the plasma. Specifically, we studied how the changes in buffer gas, EUV intensity, EUV spectrum, and secondary electron emission change the  plasma in the cavity.

We found that the EUV-induced plasma can be in a distinct ``charged'' regime. In this regime (as opposed to the quasi neutral regime, that corresponds to usual plasmas, such as glow discharges) the quasi neutrality is violated everywhere in the plasma. This regime forms due to the escape of fast photoelectrons from the  plasma to the cavity walls. That leaves a highly charge imbalanced plasma. Such a plasma is not well suited for study with the microwave cavity method. However, it happens that such a regime is on the lower boundary of the methods sensitivity. Therefore, it can be easily detected and avoided. 

We evaluated the accuracy of the simulations by comparing of the simulated  field averaged electron density with the experimentally measured electron densities. We show that in the quasi-neutral plasma regime, the microwave cavity resonance shift provides an accurate tool for estimating plasma parameters, using only the spatial profile of the cavity mode and the EUV beam profile as input factors. In the charged regime, similar accuracy can only be obtained if on has knowledge of the EUV spectrum and secondary electron emission among other factors.

\section {Experiment \label{section:experiment}}
The experimental setup is described in detail in Ref.~\cite{Horst.2014.exploring}. Therefore, here, we provide a minimal outline of the experimental procedure. 

\begin{figure}
    \includegraphics[width=0.7\columnwidth]{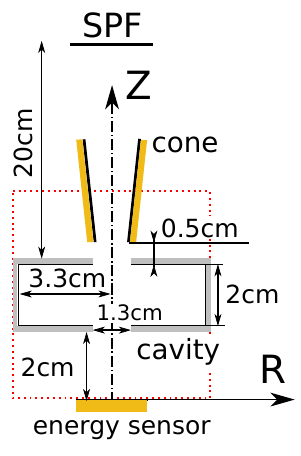}
    \caption{Configuration of the experimental chamber.              
            The EUV radiation is introduced via a spectral purity filter (SPF) along the symmetry axis of the cavity. 
            \label{fig:mw_cavity}
    }
\end{figure}

The plasma is ignited in a resonant cavity (see Fig.~\ref{fig:mw_cavity} ) by a beam of EUV radiation, which is introduced  along the axis of the cylindrically symmetric  cavity with an aperture and beam guide to prevent residual EUV from being incident on the cavity walls. The EUV radiation is produced by a xenon based discharge source  and propagated through a spectral purity filter (SPF) to limit the spectrum to photon energies to the range of 70 to 120~eV. 
The spectrum of the source after transmission through the SPF  is shown in Fig.~\ref{fig:Xe_source_spectrum}. 

The EUV power is monitored using a sensitive thermocouple, attached to a copper disk, located approximately 2~cm behind the cavity.

As described earlier, the average electron density in the resonant cavity was determined by measuring the resonant frequency of the microwave cavity (see \cite{Horst.2014.exploring} for details).

\section {Model}
The dynamics of the EUV-induced plasma was simulated using a two dimensional ($rz$ cylindrically symmetric) particle-in-cell plasma model with Monte Carlo collisions (PIC-MC). Both electrons and ions are represented by particles to describe  the  dynamics of the plasma accurately. The model follows the general PIC scheme, described elsewhere~\cite{Birdsall.1985.plasma}.

\subsection{Photoionization}
The absorption of EUV radiation in the volume is included as a source of slow ions and fast electrons. To simulate  photoionization, we use the measured spatial (axially averaged) and temporal profiles of the EUV pulse (see Fig.~\ref{fig:5Pa_spatial_profile} and Fig.~\ref{fig:charge_balance_in_plasma}).
We assume that the spatial profile and spectrum of the EUV source  does not vary during EUV pulse.

The spatial and temporal coordinates of electron-ion pairs  that are created by ionization events are  added to the simulation domain using a probability distribution function that is weighted by the   spatial and temporal profiles of the EUV pulse. The energy of the injected electrons is set to the difference between the photon energy, which is selected according to the EUV spectrum, and the ionization potential of the gas species (H$_2$ or Ar). The  the power spectral density of the  EUV beam is recalculated along the $z$ axis to take into account  absorption by the gas between SPF and cavity entrance  using  the Lambert--Beer law.

The electrons are emitted preferentially in the $r$ plane, with a  distribution function
\begin{equation}
      P(\theta) \sim 1 - \beta \frac{1}{4}(3 \cos^2 \theta - 1)
\end{equation}
where $\theta$ is the angle between electron velocity and EUV beam direction and $\beta$ is the anisotropy factor. This form of angular distribution corresponds to photoionization by unpolarized light~\cite{Gallagher.1987.absolute}. For hydrogen we use $\beta = 2 $, as this agrees with experiments in the energy range of the measurements. For argon, we  use data from \cite{Houlgate.1976.angular,Taylor.1977.photoelectron} for photon energies below 95~eV, and $\beta=1.3$ above 95~eV. 

The energy dependent photoionization cross-sections are taken from \cite{Chung.1993.dissociative} and \cite{Samson.1994.total} for H$_2$. 
For argon there are several measurements (e.g. \cite{Samson.1991.recent}, \cite{Chan.1992.absolute} and \cite{West.1976.absolute}). We followed the recommendation of \cite{Berkowitz.2001.atomic} and used data from \cite{Samson.1991.recent} for the total absorption cross-section. The cross-section for photoionization to Ar$^{++}$ is taken from \cite{Holland.1979.multiple}.

\subsection{EUV the spectrum \label{section:Xe_spectrum}}
To simulate the EUV plasma ignition, the spectrum of the EUV radiation must be taken into account. Unfortunately there is no simple experimental method to control the radiation spectrum in the relevant broad range of VUV to EUV photon energies (15.4~--~120~eV). 

\begin{figure}
    \includegraphics[width=0.5\textwidth]{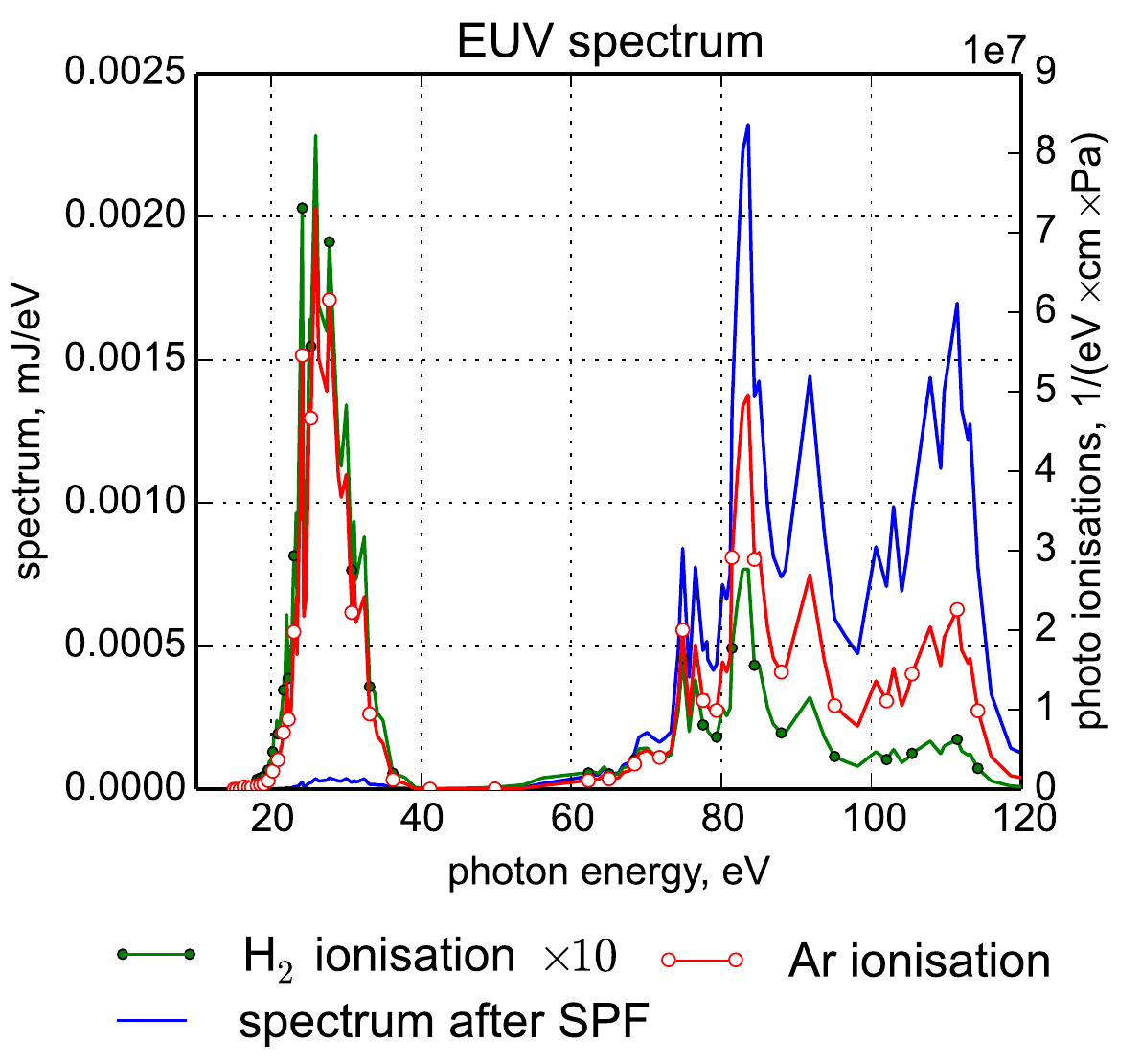}
    \caption{EUV power spectral density ($I(E)$)spectrum  as used in the simulations. The right axis shows the  number of direct photoionization events due to transmission of 0.044~mJ EUV through 1~Pa of gas, i.e. $I(E) \cdot N[\mbox{1Pa}] \cdot \sigma(E) / E$, where, $\sigma(E)$ is the photoabsorption cross-section, $E$, the photon energy, and $N[1Pa]$ the number density  . \label{fig:Xe_source_spectrum}}
\end{figure}

Therefore, as input for the simulations, we have used the source spectrum, measured before transmission through the SPF, convolved with the measured transmission spectrum of the Si:Zr SPF. This results in the EUV spectrum shown in 
Fig.~\ref{fig:Xe_source_spectrum}. 

Fig.~\ref{fig:Xe_source_spectrum} shows that the transmission of the SPF filter  in the 20~-~40~eV range, which is in the range of  1\% of the  total pulse energy, results in significant additional photoionization. This additional photoionisation  is comparable to the direct photoionization due to the high energy part of the spectrum (i.e. 60~--~120~eV), because the photo absorption cross-section is very large in the 20~-~40~eV range compared to the 60~--~120~eV range. 

Although the presence of the radiation in the 20~--~40~eV range has been confirmed by measurements, the absolute accuracy of the transmitted spectrum is not accurately known. This introduces an additional uncertainty into  the simulations. Because the accuracy of the measurements of the EUV dose per pulse is also not know, we choose to keep spectrum shape constant  and vary the EUV dose. This approach has the advantage that the   integral accuracy of the simulations can be estimated by the difference between  the measured EUV energy per pulse and that required for good agreement between experiments and simulations for both gases (i.e., we require that the same EUV power results in agreement between model and experiment over a range of experimental conditions).

\subsection{Secondary electron emission and influence of chamber configuration \label{section:chamber_configuration}} 

Emission of  slow  electrons from the walls to the plasma due to any reason   can significantly decrease the plasma potential, or even lead to the  collapse of the plasma sheath \cite{Hobbs.1967.heat} leading to effective energy transfer from the  electrons to the walls.  Therefore,  it is important to take into account secondary electron emission (SEE) from the the surfaces that are exposed to the plasma.

For the considered setup,  energy is only injected into the plasma  during the EUV pulse via photoionization and photoelectrons emission from irradiated surfaces. Therefore, the decrease of the plasma potential or sheath collapsed due to SEE can result in significant loss of fast electrons to the walls. Hence, the maximum number of ionization events also decreases.

\subsubsection{Secondary electron emission \label{section:see}}
In order to estimate the role of emission from the aluminum cavity walls due to electron impact,  we have considered three cases. The first corresponds to no secondary electron or ion induced emission from the walls. In the second case, electron emission due to both electron and ion impact is taken into account. The data for aluminum shows a significant spread of possible emission yields. Nevertheless, for the energy range of primary photoelectrons (e.g. 50~--~75~eV) the reported values are typically larger than 0.3 \cite{Baglin.2000.secondary,Lin.2005.new} so we use the yield reported for clean aluminum in Ref.~\cite{Lin.2005.new}.

The yield also significantly depends on the surface conditions. For  aluminum with a native oxide, the emission yield~\cite{Belhaj.2013.electron} is significantly higher than for pure aluminum and corresponds to the yields found for dielectric materials~\cite{Dawson.1966.secondary,Tondu.2011.electron-emission}. Therefore, in the third case, the yield for aluminum with a native oxide is used. The yield is estimated from Ref.~\cite{Belhaj.2013.electron}  for energies below 25~eV, and extrapolated linearly for  energies  above 25~eV.  

We do not discriminate between  backscattered electrons and true secondary electrons, because, for our study, only the total flux from the surface is important. In the model, we assume that most of the emitted electrons are cold, i.e. they have a significantly lower energy compared to the energy of the impacting electrons. However, for electron with energies below 10~eV this approximation is not valid, due to the higher probability of elastic backscattering. But, these backscattered electrons do not have sufficient energy to ionize more of the background gas, if not accelerated by a plasma potential. Therefore, the additional error due to this approximation is expected to be  small.

\subsubsection{influence of chamber configuration}
The configuration of the experimental setup is presented in Fig.~\ref{fig:mw_cavity}. The SPF filter, copper cone and copper disc of energy sensors emit secondary electrons under EUV irradiation. But, these electrons do not contribute directly to the ionization inside cavity, because the number of high energy secondaries is small and the electron flux is not focused. 

To estimate the influence of this effect, we have simulated  two geometrical configurations: a large volume, which includes the cone and disc (the dotted rectangle in Fig.~\ref{fig:mw_cavity}), and a  small volume, which includes only the  cavity with periodic boundary conditions at the cavity  openings. We have not included the full length of the cone and SPF into simulations, because the aspect ratio of the cone diameter to length is very small ( i.e. 0.05). 

\subsubsection{Photo electron emission from surface}
In order to model the configuration that includes the  cone and copper disc,  photoelectron emission should be taken into account. The energy spectrum of photoelectrons emitted from the copper disk and cone surfaces are calculated from~\cite{Henke.1977.electron_emission}: 
\begin{equation}
P(E) \sim \frac{E}{(E + W)^4} 
\label{eqn:se_spectrum}   
\end{equation}
Here $P(E)$ is the probability of emitting an electron with energy, $E$, from a surface with a work function, $W$. For copper  $W=5$~eV~\cite{Lide.2003.crc}. For electrons emitted from the surface, we assume an angular dependence given by a cosine emission law~\cite{Henke.1977.electron_emission}. The effective photoelectron yield from copper was set to 0.05 electron/photon~\cite{Day.1981.photoelectric}. 

\subsection{Grid resolution}
The grid resolution was chosen to resolve the Debye length. To estimate the minimum Debye radius in simulations, we assumed a temperature of 0.5~eV and an estimated  maximum plasma density. The maximum plasma density was estimated from  direct photoionization (i.e., Fig.~\ref{fig:Xe_source_spectrum}), increased by the maximum possible number of impact ionizations that the primary electrons could generate, i.e., for a 90~eV  photon, the photoelectron has an energy of  74.6~eV, therefore, the maximum number of ions produced by that electron is 5, increasing the plasma density by a factor of 5. Note, however, that, in the  simulation, the contribution of fast photoelectrons is significantly smaller due to the  contribution of inelastic collisions, and that many photoelectrons escape to the wall before generating the maximum number of ions. 

\subsection{Cross-sections sets}
Two independent cross-section sets are used for modelling hydrogen and argon. Both sets consists of electron and ion collisions with corresponding neutrals. The collisions between plasma species and three body processes are neglected due to their low probability under the conditions considered here. 

We make use of the procedure described in \cite{Nanbu.1994.simple} to perform Monte Carlo collisions with the background gas. We tested the consistency of our implementation by modeling swarm experiments and found good agreement with experimental values~\cite{Dutton.1975.survey} for the first Townsend electron ionization coefficient, the electron mobility, for H$^+$  and  H$_3^+$ mobility in hydrogen~\cite{Graham.1973.mobilities}, and for Ar$^+$ and Ar$^{++}$ mobility in argon~\cite{Hornbeck.1951.drift,Johnsen.1978.mobilities}.

\subsubsection{Hydrogen}
To accurately model electron collision related processes in hydrogen discharges with a  Monte Carlo (MC) model, one needs to take into account the differential cross-sections for ionization and excitation processes. As described in detail in \cite{Mokrov.2008.monte_carlo}, the particular choice of the angular dependence of cross-sections significantly influences the simulation results. 

For electron -- H$_2$ collisions we adopt a set of cross sections found in Ref.~\cite{Mokrov.2008.monte_carlo} with small corrections. We use an experimentally determined doubly differential cross-section for the electron impact ionization of hydrogen~\cite{Shyn.1981.doubly,Rudd.1993.doubly}. Electron elastic scattering and hydrogen electronic excitations, and angular scattering data taken from \cite{Brunger.2002.electronmolecule}.

The set of cross-sections for collisions between ions and hydrogen is based on~\cite{Simko.1997.transport} because this set provides good agreement with swarm data for ions in hydrogen. 

We neglect the formation of H$^-$, because the cross-section of dissociative electron attachment is very low, and the density of vibrationally excited hydrogen molecules too low to  significantly contribute to the production of H$^-$.

\subsubsection{Argon}
For electron -- Ar collisions we adopt a set of cross-sections found in Ref.~\cite{Hayashi.2003.bibliography,Phelps.webpage}. We add electron impact ionization of Ar to  Ar$^{++}$ from  Ref.~\cite{Rejoub.2002.determination} to the cross-sections set. We use differential cross-section data for electron impact ionization from Ref.~\cite{Yates.2011.near-threshold}. The data from Ref.~\cite{Yates.2011.near-threshold} covers incident electron energy range 17~--~30~eV. To the authors knowledge there is no systematic differential ionization cross-section measurements for incident electrons in the energy range of 30~--~100~eV.  Hence, for energy range above 30~eV, we use empirical formulas from Ref.~\cite{Fiala.1994.two-dimensional}. 

We neglect ionization via the metastable argon exited state. The density of metastables produced in one EUV pulse is comparable to the plasma density (e.g. $\sim 10^9~-~10^{10}$~1/cm$^3$). The maximum of cross-section for electron impact ionization of the metastable argon exited state is approximately $10^{-15}$~1/cm$^2$~\cite{Deutsch.1999.calculated}. Therefore, the maximum corresponding ionization frequency can estimated as $\nu_i = n\sigma v \sim 2\cdot 10^3$~1/s. The simulated plasma evolution is less than 10~$\mu s$, yielding an upper bound of 0.02 ionization events per electron for the step wise ionization. This is negligible compared to the direct EUV ionization and ionization by fast photoelectrons. Moreover, there is no pulse-to-pulse accumulation because the time between EUV pulses is larger than metastable's lifetime due to quenching on the chamber walls.

The set of cross-sections for collisions between Ar$^+$ and argon is based on~\cite{Phelps.1994.application}. For collisions between Ar$^{++}$ and argon we use data from \cite{Okuno.1986.charge}.

\subsection{Calculation of the field average electron density}
In order to compare the model results with experimental measurements, the simulated spatially-resolved electron density was averaged over the spatial profile of the resonant microwave cavity mode as follows.
\begin{equation}
    \langle n_e\rangle = \frac{\int n_e \vec{E}^2 dV}{\int \vec{E}^2 dV} \label{eqn:field_averaging}
\end{equation}
The magnitude of the microwave field was very small (mV range), therefore, it was not included in PIC-MC simulations.

\section {Results and discussion}
\begin{figure} 
  \includegraphics[width=\columnwidth]{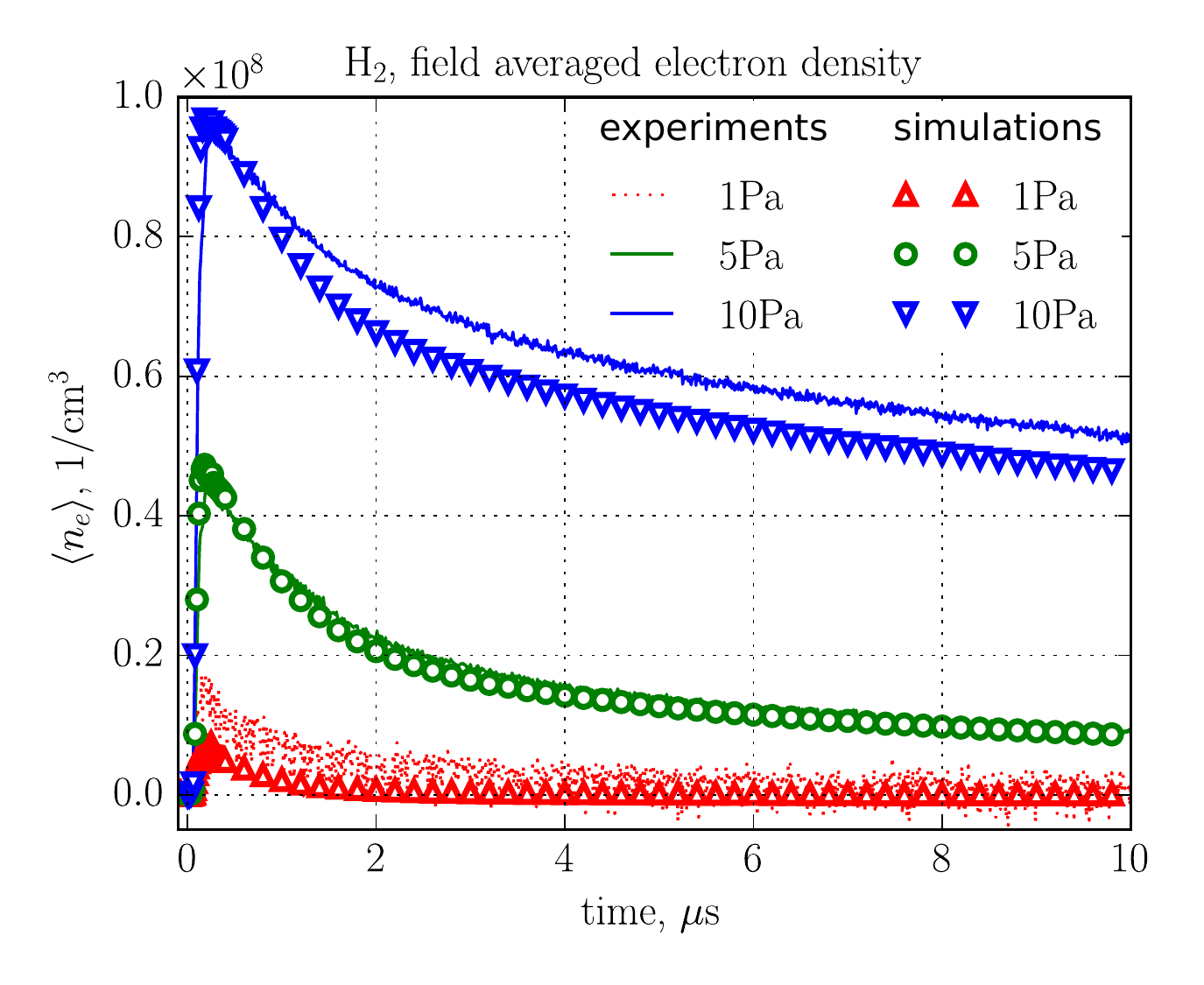}   
  \caption{Comparison between the simulated (symbols) and measured (lines, data from part~I) time dependence of field averaged electron densityfor 1 (red), 5 (green), and 10 (blue)~Pa hydrogen. \label{fig:averaged_ne_h2}}
\end{figure}

\begin{figure}   
      \includegraphics[width=\columnwidth]{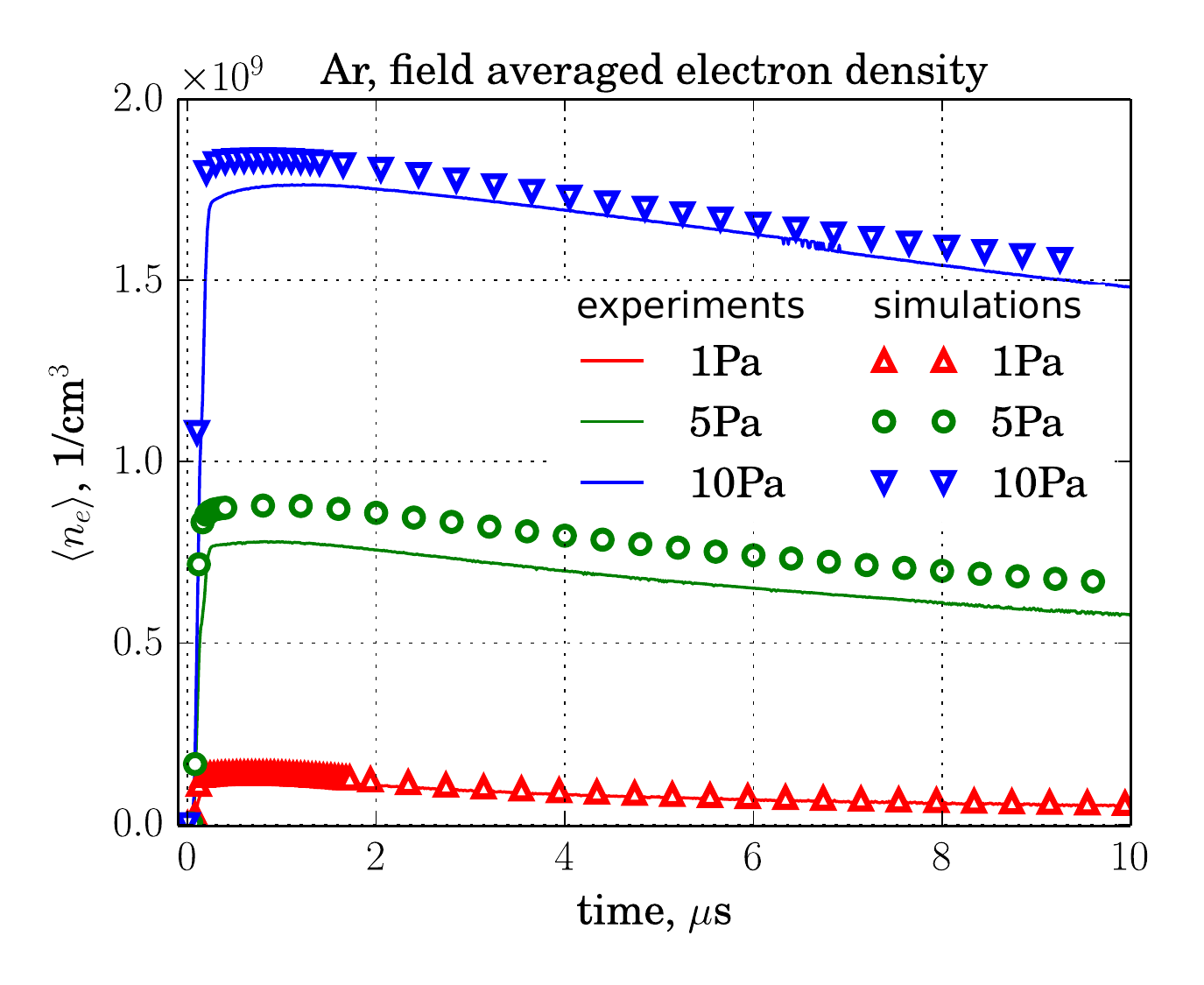} 
   \caption{Comparison between the simulated (symbols) and measured (lines, data from~\cite{Horst.2015.exploring}) time dependence of field averaged electron density for 1 (red), 5 (green), and 10 (blue)~Pa argon.  \label{fig:averaged_ne_Ar}}
\end{figure}

The ignition  and decay of the EUV induced plasma was simulated for a  series of  argon and hydrogen pressures (i.e. 1~Pa, 5~Pa, 10~Pa) to compare with experimental results. The best agreement with experiments for both gases and all considered pressures was obtained in simulations where we have included the larger chamber configuration, and used a spectrum that includes a VUV contribution. The comparison between simulated field averaged electron density and experimental measurements is presented in Fig.~\ref{fig:averaged_ne_h2} and Fig.~\ref{fig:averaged_ne_Ar}. 

In these simulations, we have decreased the EUV energy per pulse to 0.035~mJ/pulse from the experimentally measured value of 0.044~mJ/pulse. This difference, in combination with $\sim 20\%$ relative errors for the field averaged electron densities (i.e. in Fig.~\ref{fig:averaged_ne_h2} and Fig.~\ref{fig:averaged_ne_Ar}), yields a $\sim 50\%$ cumulative uncertainty, which is comparable to the $\sim$30\% error margin of  of the experiment~\cite{Horst.2014.exploring}.

\subsection{Relation of field averaged electron density and simulated values \label{section:mw_mesured_density_and_sim}}

\begin{figure}
    \includegraphics[width=\columnwidth]{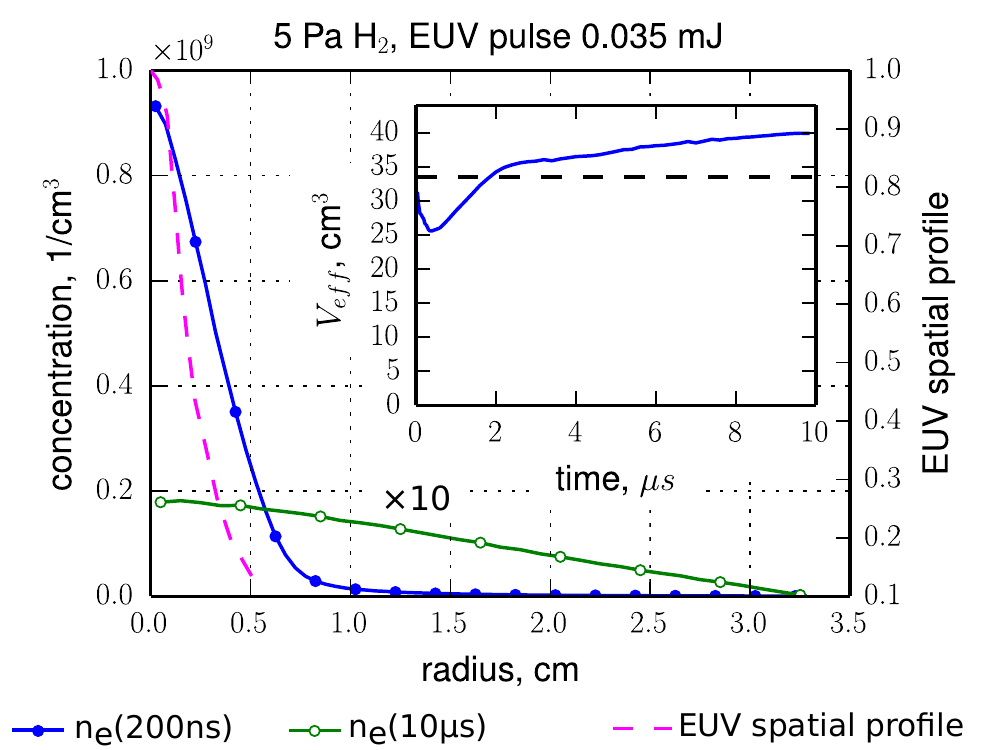} 
   \caption{    
    Radial distributions of electron density in the cavity for 5~Pa H$_2$ and the EUV spatial profile that was used for all simulations. The insert shows the  time dependence of the effective volume probed by the microwave mode. The simulated volume is a  solid blue line, while the dashed black line corresponds to analytical estimation (see text).  \label{fig:5Pa_spatial_profile}}
\end{figure}

The cavity method allows the field averaged electron density to be determined. But, the relationship of this quantity to the actual plasma density is not straightforward. During the first 1~$\mu s$, the simulated plasma is much denser than that measured by the cavity. However, as the plasma expands, the difference between simulated plasma density and cavity measurements becomes smaller. For 5~Pa H$_2$ at 10~$\mu$s, the ratio between the simulated plasma density and that measured by the cavity is only about a factor of two (see Fig.~\ref{fig:5Pa_spatial_profile}).

Nevertheless, the plasma density measured  by the cavity is very useful to estimate the amount  of the plasma in the cavity.
Let us define the effective volume of the cavity as ratio of the full number of electrons in the cavity to the cavity measured electron density i.e. $ V_{eff} =  \int n_e dV /\langle n_e \rangle $.  In simulations, this quantity varies only by about a factor of two during the 10~$\mu s$ simulation window (see insert in Fig.~\ref{fig:5Pa_spatial_profile}). 
Moreover, the estimation of the effective volume, based on the EUV spatial profile (as an approximation to the EUV plasma distribution), and the spatial profile of the microwave cavity mode (see Part I) agree well with the simulated values (dashed black line in the insert in  Fig.~\ref{fig:5Pa_spatial_profile})  

Therefore, the amount of EUV induced plasma can be estimated from the measured cavity response with a relative accuracy of about 30\%. Hence, the amount of absorbed radiation can be estimated with similar accuracy, if the other factors (see below) that effect plasma formation are eliminated. 

The simulated evolution of the effective volume is non monotonic in time. A similar effect was observed in experiments via the ratio of the electron densities measured using 010 and 110 cavity modes~\cite{vanderHorst.2015.dynamics}. However, in simulations, the decrease of the effective volume just after the EUV pulse is caused by the convolution of the plasma density with the spatial profile of the microwave mode in the cavity.

\subsection{Effect of VUV part of the spectrum}
As discussed in section~\ref{section:Xe_spectrum}, the EUV spectrum after passing through the SPF (see Fig.~\ref{fig:Xe_source_spectrum}) has two distinct regions: a low energy (20~--~40~eV) VUV part, and a high energy (60~--~120~eV) EUV part. Our simulations show that the VUV contribution  is needed to produce consistent results for both gases with the same amount of the radiation dose per pulse. This is because the average number of ionization events  per  absorbed photon can be larger than one if the emitted fast photoelectron produces addition ionization events. 

In our simulations, the average number of ionizations per absorbed EUV photon was observed to be approximately one for H$_2$ and two for Ar. To obtain the observed electron densities without including ionization due to  VUV, the EUV pulse energy must be increased and set to different values for argon and hydrogen.

\subsection{Effect of secondary electron emission due to EUV radiation}

\begin{figure}
    \includegraphics[width=0.45\textwidth]{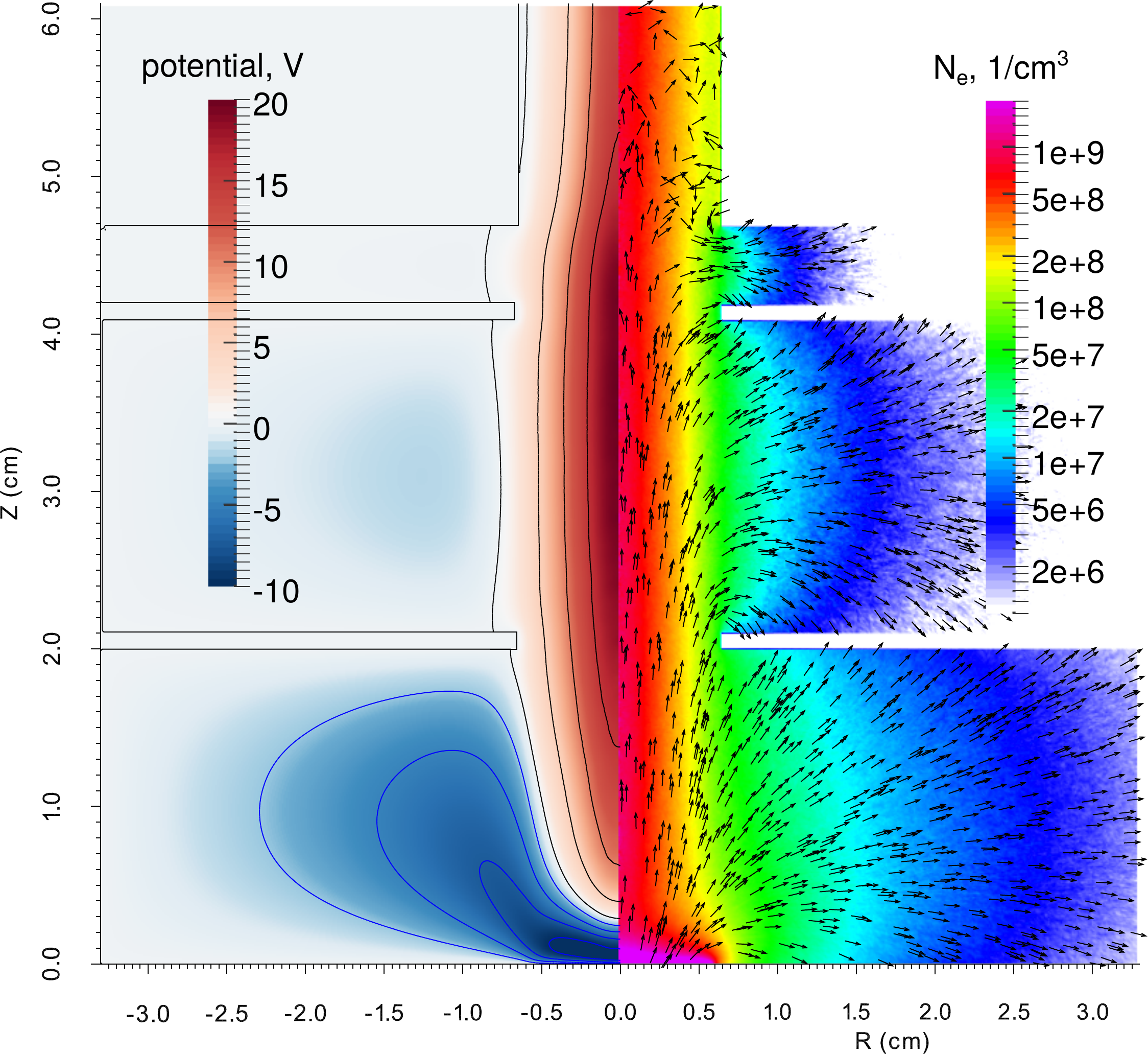}
    \caption{Distribution of the potential, electron density and current direction in the simulation domain 100~ns (see Fig.~\ref{fig:charge_balance_in_plasma}), 5~Pa H$_2$. The arrows shows the direction of the electron current. The electrons are drained from the copper disc (Z $\sim$0) into the cavity, leading to a decrease of the plasma potential (see Fig.~\ref{fig:geometry_effect}).  \label{fig:plasma_potential_2D} }
\end{figure}

\begin{figure}
    \includegraphics[width=0.45\textwidth]{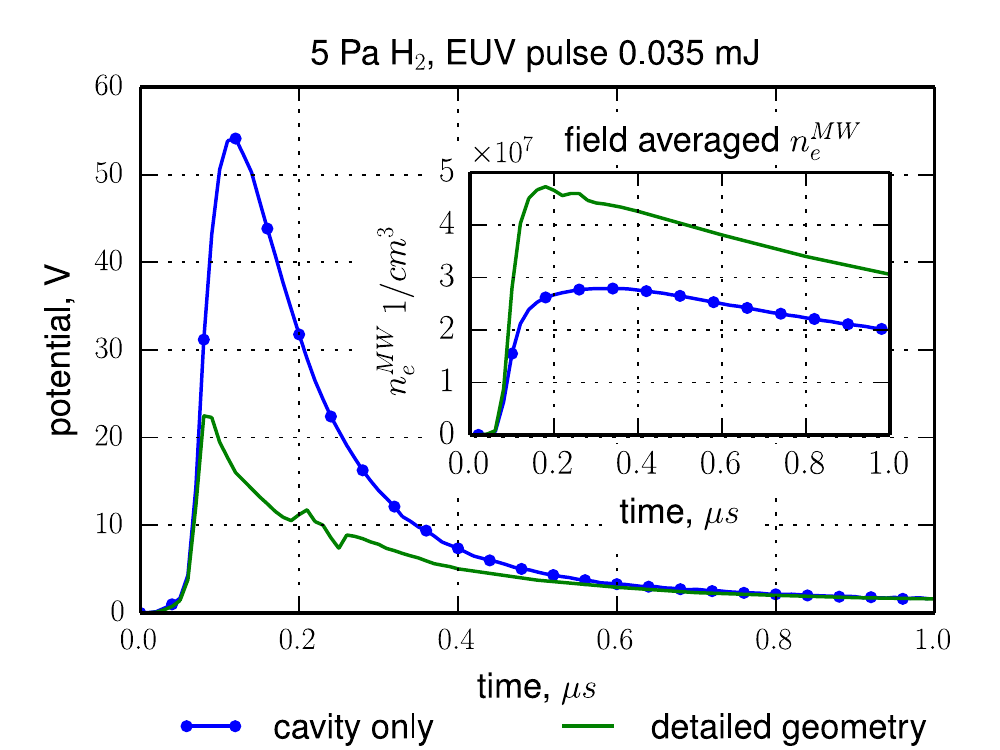}
    \caption{Comparison of the potential in the center of the cavity as function of time for 5~Pa H$_2$. The large geometrical configuration includes secondary electron emission from the energy sensor, which leads to a decrease of the potential in the cavity. A comparison between the field averaged electron densities in the cavity is shown in the insert.
    A non-monotonic decrease of the plasma potential for the large configuration is related to SE emission from energy sensor and charge redistribution in the cavity. During the EUV pulse, SE emission leads to a significant decrease of the plasma potential, since the emitted SE electrons replace the escaped primary photoelectrons. Just after the EUV pulse there are no SEs emitted, but some fast photoelectrons are still present in the plasma, therefore, once these fast electrons escape to the walls the plasma potential increases temporally. \label{fig:geometry_effect} }
\end{figure}

Our simulations of the larger, more complete geometry shows that electron emission from the cone walls and copper disk change of the plasma potential. The large EUV induced secondary  electron emission from the copper disk leads to the formation of the space charge potential well during EUV pulse, see Fig.~\ref{fig:plasma_potential_2D}. The electron current during the pulse on the symmetry axis of the system is directed from the copper disk towards the cavity, thus replacing a large portion of the fast electrons that reach cavity walls. 
A comparison between  the potential in the cavity for 5~Pa H$_2$ for both geometrical configurations is presented in Fig.~\ref{fig:geometry_effect}.  For these simulations, the electron induced electron emission from cavity walls was set to zero.

Interestingly,  the average electron density  decreases for argon gas, but  increases for hydrogen,  compared to simulations of the cavity only. In the case of hydrogen, the plasma potential is smaller than for argon, due to a lower plasma density. That, in turn, leads to a significant loss of electrons to the cavity walls. Electrons emitted from the cone and copper disc are accelerated  into the  plasma, compensating for a fraction of the escaped electrons and increase the electron density. 

In contrast to hydrogen, the argon plasma is significantly denser, due to the  order of magnitude larger absorption of EUV for the same pressure (see Fig.~\ref{fig:Xe_source_spectrum}). Hence, for argon, the relative charge imbalance between electrons and ions is much smaller than for hydrogen. Therefore, these electrons change the potential landscape of the plasma, increasing the probability of fast photoelectrons escape to the walls and reducing the average number of ionizations per fast electron.

\subsection{Influence of electron induced secondary emission from cavity walls}
As discussed in section~\ref{section:see} one can expect a significant secondary electron emission yield from the cavity walls under electron impact. However, the inclusion of this effect in the simulations has a small impact on the plasma formation in the cavity (below $5\%$ for the field averaged electron density).  
\begin{figure}
    \includegraphics[width=0.45\textwidth]{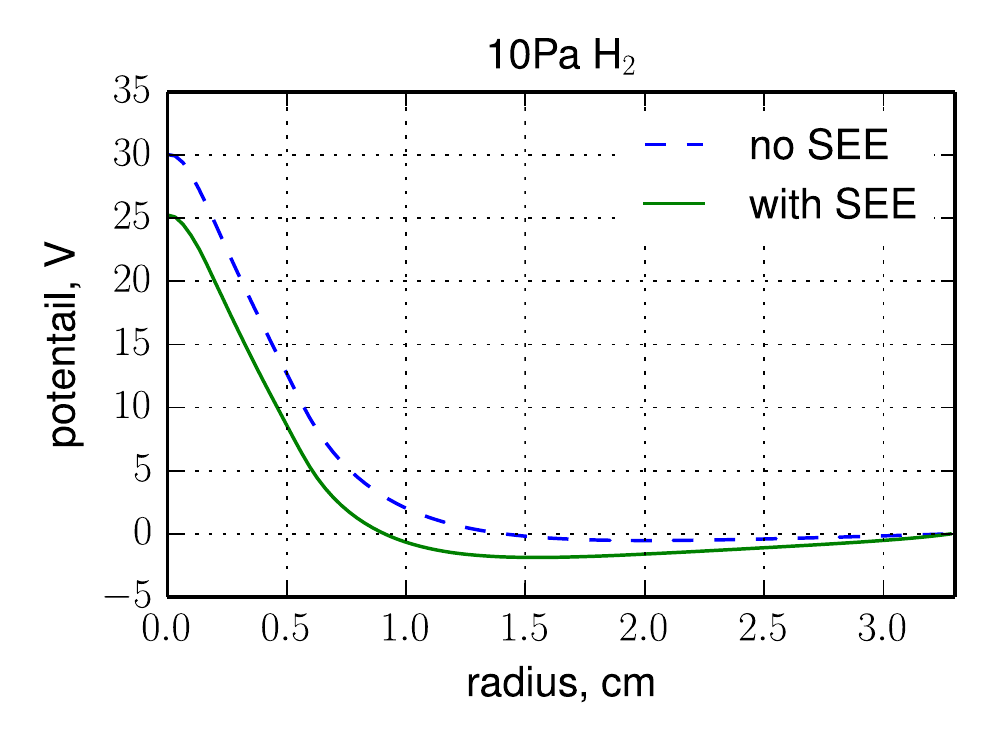}
    \caption{Comparison of the radial dependence of the plasma potential at the center of the cavity with (green) and without SEE (blue) for 10~Pa H$_2$ at 80~ns after pulse start. \label{fig:plasma_potential}}
\end{figure}

It is explained by changes in the plasma potential (see Fig.~\ref{fig:plasma_potential}). 
Most ionization occurs  in the region corresponding to the EUV beam. Electrons that escape from the plasma to the region between the beam and the cavity walls encounter the slow electrons produced on the cavity walls due to electron impact emission. That lead to formation of  a local minimum in  the plasma potential. 

Hence, the potential barrier that electrons in the beam need to overcome in order to escape is similar for both simulations that include and neglect electron induced emission from the cavity wall. Moreover, the potential near cavity wall during the pulse prevents the emitted secondary electron from entering the plasma, thus significantly limiting the influence of SEE on the plasma formation.

\subsection{High charge imbalance plasma for 1Pa H$_2$}

\begin{figure}
    \includegraphics[width=0.45\textwidth]{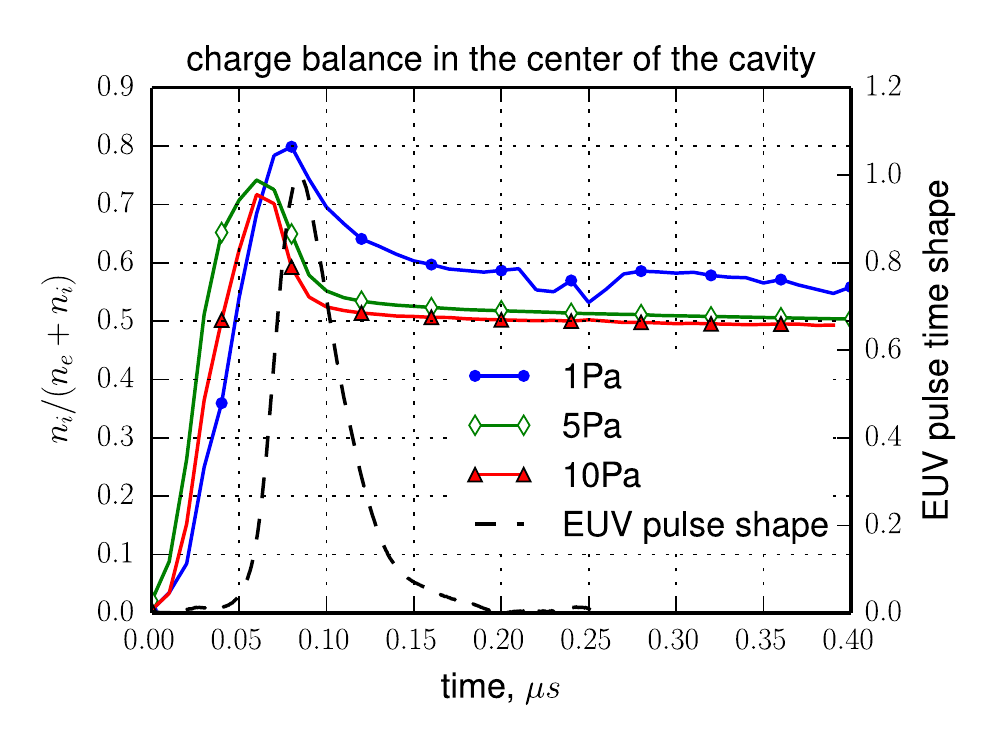}
    \caption{
      Evolution of the charge balance in the center of the cavity for the case of H$_2$. The charge imbalance for the case of 1~Pa H$_2$ lasts significantly longer compared to other cases.       
     \label{fig:charge_balance_in_plasma} }
\end{figure}

\begin{figure}
   \includegraphics[width=0.45\textwidth]{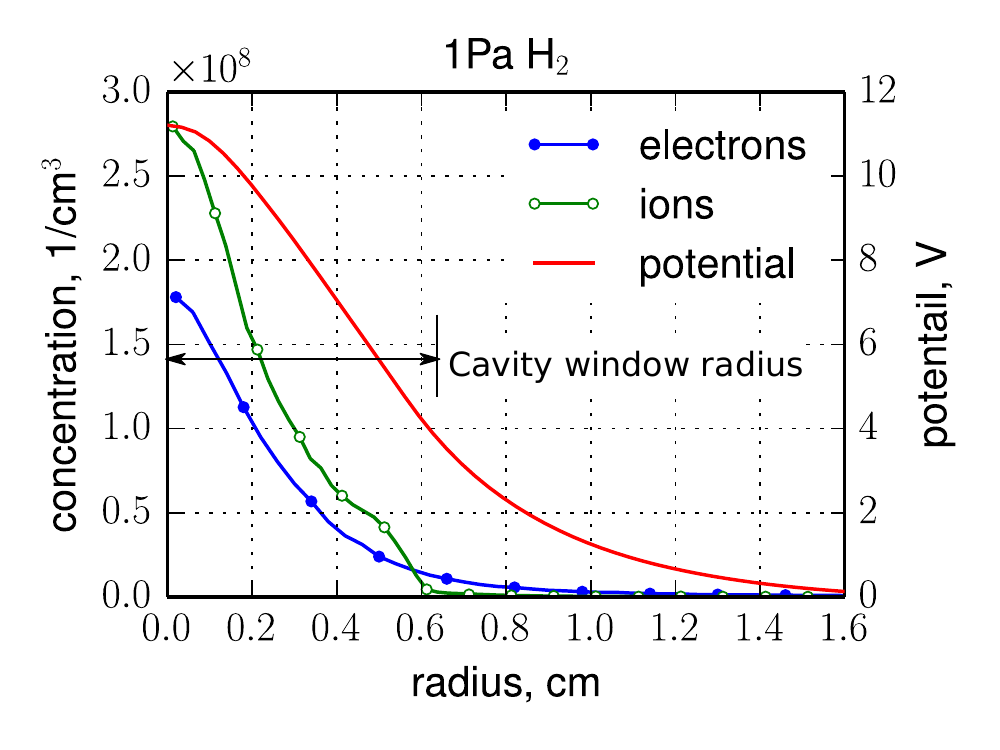}
   \caption{Radial distribution of electrons and ions in the center of the cavity at 150~ns after EUV pulse start. \label{fig:charge_balance_radial_distribution}}
\end{figure}

Interestingly, results for 1~Pa H$_2$, even with inclusion of the VUV part of spectrum, show a significant charge imbalance between ions and electron.
Due to the low H$_2$ pressure, the absorption of EUV radiation is low. The generated  plasma has a peak density in the range of $2\cdot10^8$~--~$3\cdot10^8$~1/cm$^3$, see Fig.~\ref{fig:charge_balance_radial_distribution}. This plasma can generate only a low plasma potential, that is not enough to confine high energy primary photoelectrons. As result, after the EUV pulse, a significant fraction of electrons leaves the plasma.

The ratio of the ion density to the sum of the ion and electrons densities  as a function of time are presented in Fig.~\ref{fig:charge_balance_in_plasma}. For all simulated cases, in the beginning of the pulse, many fast electrons leave the plasma, creating an excess of ions. After the pulse, the simulation show that the plasma returns to quasi-neutrality (i.e. $n_i/(n_i + n_e) \sim 0.5$), except for the case of 1~Pa H$_2$ . In the case of 1~Pa, the generated plasma remains significantly charged, which leads to a faster decay, due to Coulomb repulsion on the time scale of $1~\mu s$. 

\section{Conclusion}
We simulated the evolution of an EUV induced plasma in a microwave resonant cavity. During simulations we observed that the cavity measured electron density can be used to estimate the integral amount of ionization in the cavity. 

We show that these estimates are reasonably accurate ( $< 30\%$ for the considered conditions). This is even the case for estimates that use only the spatial mode profile of the microwave cavity mode and the EUV spatial beam profile.
Therefore, a suitable microwave cavity can be used as a valuable diagnostics tool for EUV induced plasmas.

However,  the cumulative uncertainties of the SPF transmission and the  possible error margin in the measurements of the EUV source spectrum are important for the analysis of EUV plasma formation in hydrogen and argon. This is due to the rapid increase of the absorption cross-section with decreasing photon energy. Even 1$\%$ transmission of the  SPF filter in the range of 20~--~30~eV has a significant effect on the EUV plasma formation. Moreover, the same spectral range is very important for surface chemistry processes, because the photo-absorption cross-sections are typically very large (e.g. the convolution of the cross-section with the spectrum can be similar for in-band EUV and out-of-band VUV).

The effect of VUV is likely to be significant, because most laboratory sources are discharge based with grazing incidence collector optics, which, without special care will have a significant amount of VUV. Therefore, in laboratory experiments, the EUV spectrum should be measured after the SPF filter, otherwise the remaining VUV radiation can induce a substantial systematic error.

Last but not least, it is important to take into account the EUV (or VUV) induced secondary electron emission (SE) form cavity and/or surroundings. This emission changes the potential distribution in the cavity and, hence, changes the generated plasma density, which further complicates the analysis of the cavity response.

Nevertheless, in case the radiation spectrum and secondary electron emission are known through additional measurements, the microwave cavity response may prove to be a very accurate tool for estimating plasma parameters and related phenomena.

\section*{Acknowledgements}
This work is part of the research programme ‘Controlling photon and plasma induced processes at EUV optical surfaces (CP3E)’ of the ‘Stichting voor Fundamenteel Onderzoek der Materie (FOM)’, which is financially supported by the ‘Nederlandse Organisatie voor Wetenschappelijk Onderzoek (NWO)’. The CP3E programme is co-financed by Carl Zeiss SMT and ASML. The authors also acknowledge support from the Province of Overijssel, PANalytical, DEMCON, Solmates, and the University of Twente through the programme of the Industrial Focus Group XUV Optics.

\bibliographystyle{unsrtnat}
\bibliography{refs}

\begin{thebibliography}{46}
\providecommand{\natexlab}[1]{#1}
\providecommand{\url}[1]{\texttt{#1}}
\expandafter\ifx\csname urlstyle\endcsname\relax
  \providecommand{\doi}[1]{doi: #1}\else
  \providecommand{\doi}{doi: \begingroup \urlstyle{rm}\Url}\fi

\bibitem[Madey et~al.(2006)Madey, Faradzhev, Yakshinskiy, and
  Edwards]{Madey.2006.surface}
Theodore~E. Madey, Nadir~S. Faradzhev, Boris~V. Yakshinskiy, and N.V. Edwards.
\newblock Surface phenomena related to mirror degradation in extreme
  ultraviolet ({{EUV}}) lithography.
\newblock \emph{Applied Surface Science}, 253\penalty0 (4):\penalty0
  1691--1708, 2006.
\newblock \doi{10.1016/j.apsusc.2006.04.065}.

\bibitem[Davis et~al.(2007)Davis, Kyriakou, Grant, Tikhov, and
  Lambert]{Davis.2007.situ}
David~J. Davis, Georgios Kyriakou, Robert~B. Grant, Mintcho~S. Tikhov, and
  Richard~M. Lambert.
\newblock Toward the {{In Situ Remediation}} of {{Carbon Deposition}} on
  {{Ru-Capped Multilayer Mirrors Intended}} for {{EUV
  Lithography}}:\hspace{0.167em} {{Exploiting}} the {{Electron-Induced
  Chemistry}}.
\newblock \emph{The Journal of Physical Chemistry C}, 111\penalty0
  (33):\penalty0 12165--12168, 2007.
\newblock \doi{10.1021/jp074766y}.

\bibitem[{R. M. van der Horst} et~al.(2014){R. M. van der Horst}, {J. Beckers},
  {S. Nijdam}, and {G. M. W. Kroesen}]{Horst.2014.exploring}
{R. M. van der Horst}, {J. Beckers}, {S. Nijdam}, and {G. M. W. Kroesen}.
\newblock Exploring the temporally resolved electron density evolution in
  extreme ultra-violet induced plasmas.
\newblock \emph{Journal of Physics D: Applied Physics}, 47\penalty0
  (30):\penalty0 302001, July 2014.
\newblock \doi{10.1088/0022-3727/47/30/302001}.

\bibitem[Gundermann et~al.(2001)Gundermann, Loffhagen, Wagner, and
  Winkler]{Gundermann.2001.microwave}
S.~Gundermann, D.~Loffhagen, H.-E. Wagner, and R.~Winkler.
\newblock Microwave diagnostics of the electron density in molecular mixture
  plasmas.
\newblock \emph{Contributions to Plasma Physics}, 41\penalty0 (1):\penalty0
  45--60, 2001.

\bibitem[Velden et~al.(2008)Velden, Kroesen, Mullen, Moors, and {Technische
  Universiteit Eindhoven}]{Velden.2008.radiation}
van der MHL~(Marc) Velden, GMW~(Gerrit) Kroesen, van der JJAM~(Joost) Mullen,
  JHJ~(Roel) Moors, and {Technische Universiteit Eindhoven}.
\newblock \emph{Radiation generated plasmas:a challenge in modern lithography}.
\newblock PhD thesis, Technische Universiteit Eindhoven, 2008.

\bibitem[Dolgov et~al.(2015)Dolgov, Yakushev, Abrikosov, Snegirev, Krivtsun,
  Lee, and Bijkerk]{Dolgov.2015.extreme}
A~Dolgov, O~Yakushev, A~Abrikosov, E~Snegirev, V~M Krivtsun, C~J Lee, and
  F~Bijkerk.
\newblock Extreme ultraviolet ({{EUV}}) source and ultra-high vacuum chamber
  for studying {{EUV}}-induced processes.
\newblock \emph{Plasma Sources Science and Technology}, 24\penalty0
  (3):\penalty0 035003, June 2015.
\newblock \doi{10.1088/0963-0252/24/3/035003}.

\bibitem[Birdsall and Langdon(1985)]{Birdsall.1985.plasma}
Charles~K Birdsall and A.~Bruce Langdon.
\newblock \emph{Plasma physics via computer simulation}.
\newblock {McGraw-Hill}, New York, 1985.
\newblock ISBN 0-07-005371-5 978-0-07-005371-7.

\bibitem[Gallagher et~al.(1987)Gallagher, Brion, Samson, and
  Langhoff]{Gallagher.1987.absolute}
J.W. Gallagher, C.~E. Brion, J.A.R. Samson, and P.W. Langhoff.
\newblock Absolute {{Cross Sections}} for {{Molecular Photoabsorption}},
  {{Partial Photoionization}}, and {{Ionic Photofragmentation Processes}}.
\newblock \emph{Journal of Physical and Chemical Reference Data}, 17\penalty0
  (1):\penalty0 9--153, 1987.
\newblock \doi{10.1063/1.555821}.

\bibitem[Houlgate et~al.(1976)Houlgate, West, Codling, and
  Marr]{Houlgate.1976.angular}
R.~G. Houlgate, J.~B. West, K.~Codling, and G.~V. Marr.
\newblock The angular distribution of the 3p electrons and the partial cross
  section of the 3s electrons of argon from threshold to 70 {{eV}}.
\newblock \emph{Journal of Electron Spectroscopy and Related Phenomena},
  9\penalty0 (2):\penalty0 205--209, 1976.
\newblock \doi{10.1016/0368-2048(76)81030-4}.

\bibitem[Taylor(1977)]{Taylor.1977.photoelectron}
K.~T. Taylor.
\newblock Photoelectron angular-distribution beta parameters for neon and
  argon.
\newblock \emph{Journal of Physics B: Atomic and Molecular Physics},
  10\penalty0 (18):\penalty0 L699, 1977.

\bibitem[Chung et~al.(1993)Chung, Lee, Masuoka, and
  Samson]{Chung.1993.dissociative}
Y.~M Chung, E.~M Lee, T.~Masuoka, and James A.~R Samson.
\newblock Dissociative photoionization of {{H2}} from 18 to 124 {{eV}}.
\newblock \emph{The Journal of Chemical Physics}, 99\penalty0 (2):\penalty0
  885--889, July 1993.
\newblock \doi{doi:10.1063/1.465352}.

\bibitem[Samson and Haddad(1994)]{Samson.1994.total}
James A.~R. Samson and G.~N. Haddad.
\newblock Total photoabsorption cross sections of {{H}}2 from 18 to 113 {{eV}}.
\newblock \emph{Journal of the Optical Society of America B}, 11\penalty0
  (2):\penalty0 277--279, February 1994.
\newblock \doi{10.1364/JOSAB.11.000277}.

\bibitem[Samson et~al.(1991)Samson, Lyn, Haddad, and Angel]{Samson.1991.recent}
J.~A.R. Samson, L.~Lyn, G.~N. Haddad, and G.~C. Angel.
\newblock {{RECENT PROGRESS ON THE MEASUREMENT OF ABSOLUTE ATOMIC
  PHOTOIONIZATION CROSS SECTIONS}}.
\newblock \emph{Le Journal de Physique IV}, 01\penalty0 (C1):\penalty0
  C1--99--C1--107, March 1991.
\newblock \doi{10.1051/jp4:1991113}.

\bibitem[Chan et~al.(1992)Chan, Cooper, Guo, Burton, and
  Brion]{Chan.1992.absolute}
W.~F. Chan, G.~Cooper, X.~Guo, G.~R. Burton, and C.~E. Brion.
\newblock Absolute optical oscillator strengths for the electronic excitation
  of atoms at high resolution. {{III}}. {{The}} photoabsorption of argon,
  krypton, and xenon.
\newblock \emph{Physical Review A}, 46\penalty0 (1):\penalty0 149--171, July
  1992.
\newblock \doi{10.1103/PhysRevA.46.149}.

\bibitem[West and Marr(1976)]{West.1976.absolute}
J.~B. West and G.~V. Marr.
\newblock The {{Absolute Photoionization Cross Sections}} of {{Helium}},
  {{Neon}}, {{Argon}} and {{Krypton}} in the {{Extreme Vacuum Ultraviolet
  Region}} of the {{Spectrum}}.
\newblock \emph{Proceedings of the Royal Society A: Mathematical, Physical and
  Engineering Sciences}, 349\penalty0 (1658):\penalty0 397--421, May 1976.
\newblock \doi{10.1098/rspa.1976.0081}.

\bibitem[Berkowitz(2001)]{Berkowitz.2001.atomic}
Joseph Berkowitz.
\newblock \emph{Atomic and molecular photoabsorption: absolute total cross
  sections}.
\newblock {Academic Press}, 2001.
\newblock ISBN 978-0-08-052769-7.

\bibitem[Holland et~al.(1979)Holland, Codling, Marr, and
  West]{Holland.1979.multiple}
D.~M.~P. Holland, K.~Codling, G.~V. Marr, and J.~B. West.
\newblock Multiple photoionisation in the rare gases from threshold to 280
  {{eV}}.
\newblock \emph{Journal of Physics B: Atomic and Molecular Physics},
  12\penalty0 (15):\penalty0 2465, 1979.
\newblock \doi{10.1088/0022-3700/12/15/008}.

\bibitem[Hobbs and Wesson(1967)]{Hobbs.1967.heat}
G.~D. Hobbs and J.~A. Wesson.
\newblock Heat flow through a {{Langmuir}} sheath in the presence of electron
  emission.
\newblock \emph{Plasma Physics}, 9\penalty0 (1):\penalty0 85, 1967.

\bibitem[Baglin et~al.(2000)Baglin, Bojko, Scheuerlein, Gr{\"o}bner, Taborelli,
  Henrist, and Hilleret]{Baglin.2000.secondary}
V.~Baglin, J.~Bojko, C.~Scheuerlein, Oswald Gr{\"o}bner, M.~Taborelli, Bernard
  Henrist, and No{\"e}l Hilleret.
\newblock The secondary electron yield of technical materials and its variation
  with surface treatments.
\newblock Technical Report LHC-Project-Report-433 ;
  CERN-LHC-Project-Report-433, 2000.

\bibitem[Lin and Joy(2005)]{Lin.2005.new}
Yinghong Lin and David~C. Joy.
\newblock A new examination of secondary electron yield data.
\newblock \emph{Surface and Interface Analysis}, 37\penalty0 (11):\penalty0
  895--900, November 2005.
\newblock \doi{10.1002/sia.2107}.

\bibitem[Belhaj et~al.(2013)Belhaj, Roupie, Jbara, Puech, Balcon, and
  Payan]{Belhaj.2013.electron}
M.~Belhaj, J.~Roupie, O.~Jbara, J.~Puech, N.~Balcon, and D.~Payan.
\newblock Electron emission at very low electron impact energy:
  {{Experimental}} and {{Monte-Carlo}} results.
\newblock 2013.
\newblock \doi{10.5170/CERN-2013-002.137}.

\bibitem[Dawson(1966)]{Dawson.1966.secondary}
P.~H. Dawson.
\newblock Secondary {{Electron Emission Yields}} of some {{Ceramics}}.
\newblock \emph{Journal of Applied Physics}, 37\penalty0 (9):\penalty0 3644,
  1966.
\newblock \doi{10.1063/1.1708934}.

\bibitem[Tondu et~al.(2011)Tondu, Belhaj, and
  Inguimbert]{Tondu.2011.electron-emission}
T.~Tondu, M.~Belhaj, and V.~Inguimbert.
\newblock Electron-emission yield under electron impact of ceramics used as
  channel materials in {{Hall}}-effect thrusters.
\newblock \emph{Journal of Applied Physics}, 110\penalty0 (9):\penalty0 093301,
  November 2011.
\newblock \doi{10.1063/1.3653820}.

\bibitem[Henke et~al.(1977)Henke, Smith, and
  Attwood]{Henke.1977.electron_emission}
Burton~L. Henke, Jerel~A. Smith, and David~T. Attwood.
\newblock 0.1{\textendash}10-{{keV}} x-ray-induced electron emissions from
  solids{\textemdash}{{Models}} and secondary electron measurements.
\newblock \emph{Journal of Applied Physics}, 48:\penalty0 1852, 1977.
\newblock \doi{10.1063/1.323938}.

\bibitem[Lide(2003)]{Lide.2003.crc}
David~R Lide.
\newblock \emph{{{CRC}} handbook of chemistry and physics, 2003-2004: a
  ready-reference book of chemical and physical data.}
\newblock {CRC Press}, Boca Raton, Fla., 2003.
\newblock ISBN 0-8493-0484-9 978-0-8493-0484-2.

\bibitem[Day et~al.(1981)Day, Lee, Saloman, and Nagel]{Day.1981.photoelectric}
R.~H. Day, P.~Lee, E.~B. Saloman, and D.~J. Nagel.
\newblock Photoelectric quantum efficiencies and filter window absorption
  coefficients from 20 {{eV}} to 10 {{KeV}}.
\newblock \emph{Journal of Applied Physics}, 52\penalty0 (11):\penalty0
  6965--6973, November 1981.
\newblock \doi{10.1063/1.328653}.

\bibitem[Nanbu(1994)]{Nanbu.1994.simple}
Kenichi Nanbu.
\newblock Simple {{Method}} to {{Determine Collisional Event}} in {{Monte Carlo
  Simulation}} of {{Electron-Molecule Collision}}.
\newblock \emph{Japanese Journal of Applied Physics}, 33\penalty0 (Part 1, No.
  8):\penalty0 4752--4753, August 1994.
\newblock \doi{10.1143/JJAP.33.4752}.

\bibitem[Dutton(1975)]{Dutton.1975.survey}
J.~Dutton.
\newblock A survey of electron swarm data.
\newblock \emph{Journal of Physical and Chemical Reference Data}, 4\penalty0
  (3):\penalty0 577--856, July 1975.
\newblock \doi{10.1063/1.555525}.

\bibitem[Graham et~al.(1973)Graham, James, Keever, Albritton, and
  McDaniel]{Graham.1973.mobilities}
E.~Graham, D.~R. James, W.~C. Keever, D.~L. Albritton, and E.~W. McDaniel.
\newblock Mobilities and longitudinal diffusion coefficients of mass-identified
  hydrogen ions in {{H2}} and deuterium ions in {{D2}} gas.
\newblock \emph{The Journal of Chemical Physics}, 59\penalty0 (7):\penalty0
  3477--3481, October 1973.
\newblock \doi{doi:10.1063/1.1680505}.

\bibitem[Hornbeck(1951)]{Hornbeck.1951.drift}
John Hornbeck.
\newblock The {{Drift Velocities}} of {{Molecular}} and {{Atomic Ions}} in
  {{Helium}}, {{Neon}}, and {{Argon}}.
\newblock \emph{Physical Review}, 84\penalty0 (4):\penalty0 615--620, 1951.
\newblock \doi{10.1103/PhysRev.84.615}.

\bibitem[Johnsen and Biondi(1978)]{Johnsen.1978.mobilities}
Rainer Johnsen and Manfred~A. Biondi.
\newblock Mobilities of doubly charged rare-gas ions in their parent gases.
\newblock \emph{Physical Review A}, 18\penalty0 (3):\penalty0 989--995,
  September 1978.
\newblock \doi{10.1103/PhysRevA.18.989}.

\bibitem[Mokrov and Raizer(2008)]{Mokrov.2008.monte_carlo}
M.~S. Mokrov and Yu~P. Raizer.
\newblock Monte {{Carlo}} method for finding the ionization and secondary
  emission coefficients and {{I{\textendash}V}} characteristic of a
  {{Townsend}} discharge in hydrogen.
\newblock \emph{Technical Physics}, 53\penalty0 (4):\penalty0 436--444, April
  2008.
\newblock \doi{10.1134/S1063784208040075}.

\bibitem[Shyn et~al.(1981)Shyn, Sharp, and Kim]{Shyn.1981.doubly}
T.~W. Shyn, W.~E. Sharp, and Y.-K. Kim.
\newblock Doubly differential cross sections of secondary electrons ejected
  from gases by electron impact: 25-250 {{eV}} on {{H}}2.
\newblock \emph{Physical Review A}, 24\penalty0 (1):\penalty0 79--88, July
  1981.
\newblock \doi{10.1103/PhysRevA.24.79}.

\bibitem[Rudd et~al.(1993)Rudd, Hollman, Lewis, Johnson, Porter, and
  Fagerquist]{Rudd.1993.doubly}
M.~E. Rudd, K.~W. Hollman, J.~K. Lewis, D.~L. Johnson, R.~R. Porter, and E.~L.
  Fagerquist.
\newblock Doubly differential electron-production cross sections for
  200{\textendash}1500-{{eV}} e+{{H}}2 collisions.
\newblock \emph{Physical Review A}, 47\penalty0 (3):\penalty0 1866--1873, March
  1993.
\newblock \doi{10.1103/PhysRevA.47.1866}.

\bibitem[Brunger and Buckman(2002)]{Brunger.2002.electronmolecule}
M.~J. Brunger and S.~J. Buckman.
\newblock Electron{\textendash}molecule scattering cross-sections. {{I}}.
  {{Experimental}} techniques and data for diatomic molecules.
\newblock \emph{Physics reports}, 357\penalty0 (3):\penalty0 215--458, 2002.

\bibitem[{\v S}imko et~al.(1997){\v S}imko, Marti{\v s}ovit{\v s}, Bretagne,
  and Gousset]{Simko.1997.transport}
T.~{\v S}imko, V.~Marti{\v s}ovit{\v s}, J.~Bretagne, and G.~Gousset.
\newblock Computer simulations of {{H}}+ and {{H}}3+ transport parameters in
  hydrogen drift tubes.
\newblock \emph{Physical Review E}, 56\penalty0 (5):\penalty0 5908--5919,
  November 1997.
\newblock \doi{10.1103/PhysRevE.56.5908}.

\bibitem[{Hayashi}(2003)]{Hayashi.2003.bibliography}
{Hayashi}.
\newblock Bibliography of {{Electron}} and {{Photon Cross Sections}} with
  {{Atoms}}.
\newblock 2003.

\bibitem[Phe()]{Phelps.webpage}
Arthur {{V}}. {{Phelps}}.
\newblock \url{http://jilawww.colorado.edu/~avp/}.

\bibitem[Rejoub et~al.(2002)Rejoub, Lindsay, and
  Stebbings]{Rejoub.2002.determination}
R.~Rejoub, B.~G. Lindsay, and R.~F. Stebbings.
\newblock Determination of the absolute partial and total cross sections for
  electron-impact ionization of the rare gases.
\newblock \emph{Physical Review A}, 65\penalty0 (4):\penalty0 042713, April
  2002.
\newblock \doi{10.1103/PhysRevA.65.042713}.

\bibitem[Yates and Khakoo(2011)]{Yates.2011.near-threshold}
Brent~R. Yates and Murtadha~A. Khakoo.
\newblock Near-threshold electron-impact doubly differential cross sections for
  the ionization of argon and krypton.
\newblock \emph{Physical Review A}, 83\penalty0 (4):\penalty0 042712, April
  2011.
\newblock \doi{10.1103/PhysRevA.83.042712}.

\bibitem[Fiala et~al.(1994)Fiala, Pitchford, and
  Boeuf]{Fiala.1994.two-dimensional}
A.~Fiala, L.~C. Pitchford, and J.~P. Boeuf.
\newblock Two-dimensional, hybrid model of low-pressure glow discharges.
\newblock \emph{Physical Review E}, 49\penalty0 (6):\penalty0 5607, 1994.
\newblock \doi{10.1103/PhysRevE.49.5607}.

\bibitem[Deutsch et~al.(1999)Deutsch, Becker, Matt, and
  M{\"a}rk]{Deutsch.1999.calculated}
H.~Deutsch, K.~Becker, S.~Matt, and T.~D. M{\"a}rk.
\newblock Calculated cross sections for the electron-impact ionization of
  metastable atoms.
\newblock \emph{Journal of Physics B: Atomic, Molecular and Optical Physics},
  32\penalty0 (17):\penalty0 4249, 1999.

\bibitem[Phelps(1994)]{Phelps.1994.application}
A.~V. Phelps.
\newblock The application of scattering cross sections to ion flux models in
  discharge sheaths.
\newblock \emph{Journal of Applied Physics}, 76\penalty0 (2):\penalty0 747,
  1994.
\newblock \doi{10.1063/1.357820}.

\bibitem[Okuno(1986)]{Okuno.1986.charge}
Kazuhiko Okuno.
\newblock Charge {{Transfer}} of {{Ar}}\ensuremath{^2}+ and
  {{Kr}}\ensuremath{^2}+ in {{Their Own Gases Studied}} by the {{Beam Guide
  Technique}}.
\newblock \emph{Journal of the Physical Society of Japan}, 55\penalty0
  (5):\penalty0 1504--1515, 1986.
\newblock \doi{10.1143/JPSJ.55.1504}.

\bibitem[{van der Horst} et~al.(2015{\natexlab{a}}){van der Horst}, Beckers,
  Osorio, and Banine]{Horst.2015.exploring}
R.~M. {van der Horst}, J.~Beckers, E.~A. Osorio, and V.~Y. Banine.
\newblock Exploring the electron density in plasmas induced by extreme
  ultraviolet radiation in argon.
\newblock \emph{Journal of Physics D: Applied Physics}, 48\penalty0
  (28):\penalty0 285203, 2015{\natexlab{a}}.
\newblock \doi{10.1088/0022-3727/48/28/285203}.

\bibitem[{van der Horst} et~al.(2015{\natexlab{b}}){van der Horst}, Beckers,
  Osorio, and Banine]{vanderHorst.2015.dynamics}
R~M {van der Horst}, J~Beckers, E~A Osorio, and V~Y Banine.
\newblock Dynamics of the spatial electron density distribution of
  {{EUV}}-induced plasmas.
\newblock \emph{Journal of Physics D: Applied Physics}, 48\penalty0
  (43):\penalty0 432001, November 2015{\natexlab{b}}.
\newblock \doi{10.1088/0022-3727/48/43/432001}.

\end{thebibliography}
\end{document}